\documentclass{ws-procs11x85}

\usepackage{mathrsfs}
\usepackage{multicol}
\usepackage[tbtags]{amsmath}
\newcommand{\nc}[1]{\widehat{#1}}

\makeindex
\begin{document}

\title{ASPECTS OF NONCOMMUTATIVE GAUGE THEORIES\\ AND THEIR COMMUTATIVE EQUIVALENTS\thanks{I\lowercase{nvited talk at the 11th} R\lowercase{egional} C\lowercase{onference on} M\lowercase{athematics and} M\lowercase{athematical} P\lowercase{hysics,} IPM T\lowercase{ehran,} M\lowercase{ay 2004.}}}

\author{RABIN BANERJEE}

\address{S. N. Bose National Centre for Basic Sciences, JD Block, Sector 3, Salt Lake, Kolkata 700098, India\\E-mail: rabin@bose.res.in}

\maketitle\abstract{We discuss some exact Seiberg--Witten-type maps for noncommutative electrodynamics. Their implications for anomalies in different (noncommutative and commutative) descriptions are also analysed.}

\begin{multicols}{2}

Ever since Seiberg and Witten\cite{SW} gave a map (SW map) connecting a noncommutative (NC) gauge theory with its commutative equivalent, it has found startling applications as well as connections to other branches of physics.\cite{q1} This map ensures the stability of gauge transformations in the commutative and NC descriptions. The original map was given for the potentials and the field strengths, but it was only valid up to the leading order in the NC parameter. In this paper, we review some exact results on SW-type maps involving various objects like the action, currents, anomalies, etc.

\section*{Exact SW Map for NC Electrodynamics}

The action for NC electrodynamics is given by
\begin{equation}\label{101}
\nc{S}_{\mathrm{NC}} = -\frac{1}{4}\int\!\mathrm{d}^{4}x\,\nc{F}_{\mu\nu}\star\nc{F}^{\mu\nu},
\end{equation}
where $\nc{F}_{\mu\nu}=\partial_{\mu}\nc{A}_{\nu}-\partial_{\nu}\nc{A}_{\mu}-\mathrm{i}\,\nc{A}_{\mu}\star\nc{A}_{\nu}+\mathrm{i}\,\nc{A}_{\nu}\star\nc{A}_{\mu}$ and
\begin{equation}\label{102}
(A\star B)(x) = \left.\exp\left(\frac{\mathrm{i}}{2}\theta^{\alpha\beta}\partial_{\alpha}^{x}\partial_{\beta}^{y}\right)A(x)B(y)\right|_{x=y}.
\end{equation}
To express the action in terms of its commutative equivalents, recall that the low-energy effective action on a single D$p$-brane is given by the DBI action,\cite{q2}
\begin{multline}\label{103}
S(g_{s}, g, A, B)\\ = \frac{2\pi}{g_{s}(2\pi k)^\frac{p+1}{2}}\int\!\mathrm{d}^{p+1}x\,\sqrt{-\mathrm{det}\left(g+k(F+B)\right)},
\end{multline}
where $F_{\mu\nu}=\partial_{\mu}A_{\nu}-\partial_{\nu}A_{\mu}$ and the expression involves the usual closed string variables ($g_{s},g$)

As shown in Ref.~[\ref{SW}], there is a general description interpolating between the commutative description with closed string parameters and the NC description with open string parameters. The effective action in these variables is given by
\begin{multline}\label{104}
\nc{S}(G_{s}, G, \nc{A}, \theta)\\ = \frac{2\pi}{G_{s}(2\pi k)^\frac{p+1}{2}}\int\!\mathrm{d}^{p+1}x\,\sqrt{-\mathrm{det}\left(G+k(\nc{F}+\Phi)\right)},
\end{multline}
where the connection between the open and closed string parameters is
\begin{equation}\label{105}
\begin{split}
\frac{1}{G+k\Phi}+\frac{\theta}{k} = \frac{1}{g+kB},\\
G_{s} = g_{s}\sqrt{\frac{\mathrm{det}(G+k\Phi)}{\mathrm{det}(g+kB)}}.
\end{split}
\end{equation}
For $\Phi=B$ leading to $G=g$, $G_{s}=g_{s}$, $\theta=0$, the action $\nc{S}$ reproduces the commutative description \eqref{103} while $\Phi=0$ yields the familiar NC description. Furthermore, it was shown in Ref.~[\ref{SW}] that, in the slowly-varying-field approximation, the DBI actions are independent of the choice of $\Phi$, so that
\begin{equation}\label{106}
\nc{S}(G_{s},G,\nc{A},\theta) = S(g_{s},g,A,B)+O(\partial F)
\end{equation}
Using the identities \eqref{105} in the explicit structure for $\nc{S}$, $S$ and then exploiting \eqref{106} yields
\begin{equation}\label{107}
\begin{split}
&\int\!\mathrm{d}^{p+1}x\,\sqrt{-\mathrm{det}\left(G+k(\nc{F}+\Phi)\right)}\\
&= \int\!\mathrm{d}^{p+1}x\,\sqrt{\mathrm{det}(1+F\theta)}\\
&\qquad\qquad\times\sqrt{-\mathrm{det}\left(G+k(\Phi+\boldsymbol{F})\right)}+O(\partial F),
\end{split}
\end{equation}
where
\[
\boldsymbol{F} = \frac{1}{1+F\theta}F
\]
and a matrix notation $AB=A_{\mu\alpha}B^{\alpha\mu}$ or $(AB)_{\mu\nu}=A_{\mu\alpha}{B^{\alpha}}_{\nu}$ is being used. In the zero-slope limit $k\to 0$, Eq.~\eqref{107} for $\Phi=0$ and $p=3$ defines the exact non-linear action of $\theta$-expanded NC electrodynamics \eqref{101}:
\begin{multline}\label{108}
-\frac{1}{4}\int\!\mathrm{d}^{4}x\,\nc{F}_{\mu\nu}\star\nc{F}^{\mu\nu}\\ = \frac{1}{4}\int\!\mathrm{d}^{4}x\,\sqrt{\mathrm{det}(1+F\theta)}\left(\frac{1}{1+F\theta}F\frac{1}{1+F\theta}F\right).
\end{multline}
If we introduce an effective non-symmetric ``metric'' induced by dynamical gauge fields such that
\[
g_{\mu\nu} = \eta_{\mu\nu}+(F\theta)_{\mu\nu}, \quad g^{\mu\nu} = \left(\frac{1}{\eta+F\theta}\right)^{\mu\nu}
\]
then the NC Maxwell action \eqref{108} looks like the ordinary Maxwell action coupled to the induced metric $g_{\mu\nu}$:
\begin{multline}\label{109}
-\frac{1}{4}\int\!\mathrm{d}^{4}x\,\nc{F}_{\mu\nu}\star\nc{F}^{\mu\nu}\\ = -\frac{1}{4}\int\!\mathrm{d}^{4}x\,\sqrt{-\mathrm{det}g}\,g^{\mu\alpha}g^{\beta\nu}F_{\mu\nu}F_{\alpha\beta}.
\end{multline}
This result was earlier obtained in Refs.~[\ref{Y}, \ref{BY}]. Up to $O(\theta)$, an explicit expansion reveals
\begin{multline}\label{109-c}
-\frac{1}{4}\int\!\mathrm{d}^{4}x\,\nc{F}_{\mu\nu}\star\nc{F}^{\mu\nu}\\
\shoveleft{= -\frac{1}{4}\int\!\mathrm{d}^{4}x\,\Big[F_{\mu\nu}F^{\mu\nu}}\\
{}+\theta^{\alpha\beta}\left(2F_{\mu\alpha}F_{\nu\beta}-\frac{1}{2}F_{\alpha\beta}F_{\mu\nu}\right)F^{\mu\nu}\Big]\\{}+O(\theta^2).
\end{multline}
This agrees with the result obtained by directly using the SW map\cite{SW}:
\begin{equation}\label{spe}
\nc{F}_{\mu\nu} = F_{\mu\nu}+\theta^{\alpha\beta}(F_{\mu\alpha}F_{\nu\beta}-A_{\alpha}\partial_{\beta}F_{\mu\nu})+O(\theta^2)
\end{equation}
in the NC action \eqref{101}.

\section*{Exact SW map for sources}

Let us next consider the inclusion of the sources to \eqref{101}:
\begin{equation}\label{110}
\begin{split}
\nc{S}(\nc{A},\nc{\psi}) &= -\frac{1}{4}\int\!\mathrm{d}^{4}x\,\nc{F}_{\mu\nu}\star\nc{F}^{\mu\nu}+\nc{S}_{\mathrm{M}}(\nc{\psi},\nc{A})\\
&= \nc{S}_{\mathrm{ph}}(\nc{A})+\nc{S}_{\mathrm{M}}(\nc{\psi},\nc{A}).
\end{split}
\end{equation}
Then the equation of motion for $\nc{A}_{\mu}$ is
\begin{equation*}
\frac{\delta\nc{S}_{\mathrm{ph}}}{\delta\nc{A}_{\mu}}=\nc{\mathrm{D}}_{\nu}\star\nc{F}^{\nu\mu}=-\nc{J}^{\mu},
\end{equation*}
where
\begin{equation*}
\nc{J}^{\mu}:=\left.\frac{\delta\nc{S}_{\mathrm{M}}}{\delta\nc{A}_{\mu}}\right|_{\nc{\psi}}.
\end{equation*}
It is clear that $\nc{J}^{\mu}$ transforms covariantly and is also covariantly conserved ($\nc{\mathrm{D}}_{\mu}\star\nc{J}^{\mu}=0$).

Now the action \eqref{110} is rewritten, using the SW map \eqref{108} in particular, to obtain a $\mathrm{U}(1)$ gauge-invariant action defined on commutative space:
\begin{equation*}
\left.\nc{S}\left(\nc{A},\nc{\psi}\right)\right|_{\text{SW map}}:=S_{\mathrm{ph}}(A)+S_{\mathrm{M}}(\psi,A),
\end{equation*}
where $S_{\mathrm{ph}}(A)$ equals the right-hand side of \eqref{108}. The equation of motion is
\begin{equation*}
\frac{\delta S_{\mathrm{ph}}}{\delta A_{\mu}}=-J^{\mu},
\end{equation*}
where
\begin{equation*}
J^{\mu}:=\left.\frac{\delta S_{\mathrm{M}}}{\delta A_{\mu}}\right|_{\psi}.
\end{equation*}
This $J_{\mu}$ is gauge invariant and satisfies the ordinary conservation law ($\partial_{\mu}J^{\mu}=0$).

It is now feasible to derive a relation between $\nc{J_{\mu}}$ and $J_{\mu}$ by noticing that
\begin{equation}\label{111}
\begin{split}
\nc{J}^{\mu}(x)
&= \left.\frac{\delta \nc{S}_{\mathrm{M}}}{\delta\nc{A}_{\mu}}\right|_{\nc{\psi}}\\
&= \int\!\mathrm{d}^{4}y\left[\left.\frac{\delta S_{\mathrm{M}}}{\delta A_{\nu}(y)}\right|_{\psi}\frac{\delta A_{\nu}(y)}{\delta \nc{A}_{\mu}(x)}\right.\\
&\qquad\qquad\quad{}+\left.\left.\frac{\delta S_{\mathrm{M}}}{\delta \psi_{\alpha}(y)}\right|_{A}\frac{\delta \psi_{\alpha}(y)}{\delta \nc{A}_{\mu}(x)}\right]\\
&= \int\!\mathrm{d}^{4}y\,J^{\nu}(y)\frac{\delta A_{\nu}(y)}{\delta \nc{A}_{\mu}(x)},
\end{split}
\end{equation}
where the equation of motion for $\psi_{\alpha}$ (i.e., $\delta S_{\mathrm{M}}/\delta \psi_{\alpha}=0$) has been used. This is the exact SW map for the sources. It was earlier derived in Ref.~[\ref{BLY}]. Knowing the original SW map among the potentials, it is possible to get explicit expressions for the above map. Up to $O(\theta)$ we find
\begin{equation}\label{112}
\nc{J}^{\mu} = J^{\mu}+\left(\theta FJ\right)^{\mu}+\partial_{\alpha}\left(\theta^{\alpha\beta}A_{\beta}J^{\mu}\right)+O(\theta^{2}).
\end{equation}
For higher order corrections, we refer to Ref.~[\ref{BK}]. It is easy to check that this map ensures the stability of gauge transformations. Thus inserting the ordinary gauge transformations ($\delta J_{\mu}=0$, $\delta F_{\mu\nu} = 0, \delta A_{\mu}=\partial_{\mu}\lambda$), yields
\[
\delta\nc{J}_{\mu} = \theta^{\alpha\beta}\partial_{\alpha}J_{\mu}\partial_{\beta}\lambda = \theta^{\alpha\beta}\partial_{\alpha}\nc{J}_{\mu}\partial_{\beta}\nc{\lambda}+O(\theta^{2}),
\]
which reproduces the desired covariant transformation (up to $O(\theta)$) for $\nc{J}_{\mu}$.

\section*{A Map for Anomalies}

The relation \eqref{112} may be used to provide a map for anomalies. Taking the covariant derivative on both sides of \eqref{112} yields
\begin{equation}\label{113}
\nc{\mathrm{D}}^{\mu}\star\nc{J}_{\mu} = \partial^{\mu}J_{\mu}+\theta^{\alpha\beta}\partial_{\alpha}(A_{\beta}\partial^{\mu}J_{\mu}),
\end{equation}
where
\[
\nc{\mathrm{D}}^{\mu}\star\nc{J}_{\mu} = \partial^{\mu}\nc{J}_{\mu}+\theta^{\alpha\beta}\partial_{\alpha}A^{\mu}\partial_{\beta}J_{\mu}+O(\theta^{2}).
\]
This relation is obviously compatible for conserved currents ($\partial_{\mu}J^{\mu}=0$) in the commutative description and covariantly conserved currents ($\nc{\mathrm{D}}_{\mu}\star\nc{J}^{\mu}=0$) in the NC description. In fact it serves as a non-trivial consistency check on our formalism. What happens if there is an anomaly so that the currents are no longer conserved? This situation arises if we consider axial/chiral currents instead of vector currents and take loop effects into account. We prove that the map \eqref{113} is still valid providing a connection between the ABJ anomaly $\mathscr{A}=\partial_{\mu}J^{\mu 5}=(1/16\pi^2)\varepsilon_{\mu\nu\lambda\rho}F^{\mu\nu}F^{\lambda\rho}$ in the usual case and the planar (covariant) anomaly $\nc{\mathscr{A}}=\nc{\mathrm{D}}_{\mu}\star\nc{J}^{\mu 5}=(1/16\pi^2)\varepsilon_{\mu\nu\lambda\rho}\nc{F}^{\mu\nu}\star\nc{F}^{\lambda\rho}$ in the NC case, so that
\begin{equation}\label{114}
\nc{\mathscr{A}} = \mathscr{A}+\theta^{\alpha\beta}\partial_{\alpha}(A_{\beta}\mathscr{A}),
\end{equation}
as follows from a simple extension of \eqref{113}.
Inserting $\mathscr{A}$ and using the identity\cite{q8}
\[
\theta^{\alpha\beta}F_{\alpha\beta}\varepsilon_{\mu\nu\lambda\rho}F^{\mu\nu}F^{\lambda\rho} = -4\varepsilon_{\mu\nu\lambda\rho}(F\theta F)^{\mu\nu}F^{\lambda\rho},
\]
immediately yields the SW-transformed planar anomaly ($\nc{\mathscr{A}}(\nc{F}) = \nc{\mathscr{A}}(F, A)$) obtained by using the map \eqref{spe} in $\nc{\mathscr{A}}$. This completes the analysis up to $O(\theta)$. For higher orders it was shown in [\ref{BK}] that the anomaly map obtained by this technique holds only in the slowly-varying-field approximation; i.e., where the star ($\star$) product appearing in $\nc{\mathscr{A}}$ can be ignored.

\section*{Discussions}

We conclude our papaer by discussing some other possibilities. For example, the exact SW map for scalar field is found to be:\cite{BY}
\begin{multline}
\frac{1}{2}\int\!\mathrm{d}^{4}x\,(\nc{\mathrm{D}}^{\mu}\star\nc{\phi})\star(\nc{\mathrm{D}}_{\mu}\star\nc{\phi})\\
= \frac{1}{2}\int\!\mathrm{d}^{4}x\,\sqrt{\mathrm{det}(1+F\theta)}\left(\frac{1}{1+F\theta}\frac{1}{1+\theta F}\right)^{\mu\nu}\\
\times\partial_{\mu}\phi\partial_{\nu}\phi.
\end{multline}
The extension for non-abelian $\mathrm{U}(N)$ groups is rather non-trivial. The analogue of \eqref{108} is difficult since the corresponding string based arguments are highly ambiguous. However, there is no such problem for the sources and be get\cite{BK}
\[
\nc{J}^{\mu,a}(x) = \int\!\mathrm{d}^{4}y\,J^{\nu,b}(y)\frac{\delta A_{\nu}^{b}(y)}{\delta \nc{A}_{\mu}^{a}(x)},
\]
which is the expected modification to \eqref{111}.

\section*{Acknowledgements}

This work is based on collaborations with K. Kumar, C. Lee and H. S. Yang. I thank them all, as also the organisers of the workshop for a very pleasant atmosphere, academic or otherwise.

\end{multicols}
\end{document}